# Optical Transitions and Localized Edge States in Skewed Phosphorene Nanoribbons

Sima Shekarforoush[1], Farhad Khoeini[1] and Daryoush Shiri[2]

[1]Department of Physics, University of Zanjan, P.O. Box 45195-313, Zanjan, Iran

[2]Department of Physics, Chalmers University of Technology, Göteborg, SE 41296, Sweden

**Abstract:** Using the Tight Binding (TB) parameters extracted from Density Functional Theory (DFT) and Recursive Green's Function method (RGF), it is shown that skewed-zigzag black phosphorous (phosphorene) nanoribbons obtain large and tuneable bandgap in response to vertical and transverse electric fields. Depending on the direction of the applied field the mid-gap states could possess the localized or metallic nature i.e. non-zero mid-gap density of states. Adjustability of the bandgap and optical dipole transition matrix elements are explained based on the symmetry of involved band edge states. This promises new electronic and optical devices based on phosphorene nanoribbons.

## I. Introduction

Phosphorene or orthorhombic black phosphorus mono layers gained more attention during the past few years in parallel with other 2D and quasi-2D materials such as graphene, silicene, Transition Metal Dichalcogenides (TMDCs). This increased interest in phosphorene is due to its puckered honeycomb structure and *sp³* hybridization that results in unusual electrical and optical properties. Similar to graphene, it was also prepared using mechanical exfoliation technique [1]. Interestingly it has a strain-dependent as well as layer-dependent direct band gap which varies from 1.3 to 2 eV (for 1-5 layers) [2]. The existence of this gap raises a current on/off ratio to ~$10^5$ [1], hence it can be a good candidate for making tunable photo detectors with a wide spectral range (i.e. visible to IR) [3] and ultra-low power electronics. Unlike TMDCs, with low carrier mobility of ~100 cm²/V.s, phosphorene has a high hole mobility which that can be changed from 10000 cm²/V.s to 26000

cm$^2$/V.s. Moreover, it has different values of electric polarizability for zigzag and armchair directions which originates from high in-plane anisotropy in phosphorene [4]. This property can be used to fabricate field effect transistors and optoelectronic devices like photo detectors, solar cells and Light Emitting Diodes (LED) [5]. As phosphorene is strongly anisotropic, it shows linear dichroism between orthogonal in-plane directions [5, 6]. It was shown that phosphorene has a highly anisotropic phonon dispersion which leads to an asymmetrical thermal conductivity [7]. Similar to graphene ribbons, phosphorene nanoribbons (PNRs) can be created by cutting or other methods. Recently, wide phosphorene ribbons were successfully obtained using a combination of mechanical−liquid exfoliation [8].

A number of theoretical methods were used for electronic structure calculation of phosphorene such as Density Functional Theory, the GW approximation, pseudo potential approaches and a linear combination of atomic orbitals (LCAO) or the tight-binding method. The band structure of monolayer phosphorene was first investigated in 1981 by Takao et al. using the TB method [9]. An improved TB model based on the Slater–Koster–Harrison method was later presented by Osada [10].

The electronic and transport properties of PNRs are strongly dependent on the nanoribbon width and the chirality of the edges. The electronic properties normal and skewed phosphorene nanoribbons (nPNRs and sPNRs) with different edge geometries were also investigated in Refs. [11, 12]. Besides, the presence of a quasi-flat edge band close to the Fermi level in PNRs were demonstrated in [10, 13-14].

These quasi-localized states form at the finite system and decay exponentially with distance from edges. The nature of the edge states depends on the ribbon width and the chirality of the edges. Furthermore, they have a pronounced splitting in their energy levels. It was shown that quasi flat band for zigzag-edge and cliff-edge normal metallic nanoribbons and armchair-edge normal semiconducting nanoribbons is doubly degenerate [12], and for the zigzag-cliff ones it is non degenerate. In the nanoribbons with cliff-cliff edge geometry there is no flat bands [15]. However recent studies [11] showed that the skewed-zigzag (sZZ) and skewed-armchair (sAC) nanoribbons have semiconducting and metallic characteristics, respectively. The TB parameters in these studies were based on GW approximation for monolayer phosphorene in two models i.e. with 5 [16] and other with 10 [17] hopping parameters. It was shown that zigzag Phosphorene Nanoribbons (zPNRs) can be deformed using mechanical uniaxial compression and they become an indirect-gap semiconductor, a semimetal, a metal, and ultimately a plane square lattice form by increasing the strain value [1, 14, 18].

Another striking feature of monolayer phosphorene is its super flexibility. For example, Wei et al. showed that the tensile strain can be applied up to 27% and 30% along zigzag and armchair directions, respectively [19]. The puckered structure of phosphorene leads to negative Poisson's ratio which makes it exceptional among 2D materials [20].

In this work, we propose electric field as a key parameter to tune the width of bandgap and the nature of sub bands in order to alter the optical dipole transition matrix elements for different photon polarizations. This in turn leads to adjustable photon absorption and emission spectrum. This is of high value for nano-optical detectors in which the electric field (back gate voltage) can be used to make the absorption sensitive to the polarity and wavelength. Our calculations were performed based on simplified TB model found in [17] for $3p_z$ orbitals of sZZ nanoribbons.

The rest of the paper is organized as follows. Section II deals with the computational methods adopted for this study i.e. methods of generating band structure, the Density of States (DOS) and optical transition matrix elements. In Section III the main results which are effect of transverse and vertical electric fields on the band structure and symmetry of sub bands will be discussed in detail. In Section IV, we speculate on the possible applications of the observed effects in transistors and the electrically tuneable optical dipole transition matrix elements for photo detectors.

## II.    Model and Method

In this study, we suppose the skewed phosphorene ribbon is sandwiched between two semi-infinite ribbons as shown in figure 1b. The size of the middle region is characterized by *M* unit cells along the x-axis and $N_{sZZ}$ atoms inside a unit cell.

Schematic diagrams of the sandwiched spacer with the size of $N_{sZZ}$ =16 atoms and M = 3 unit cells is shown in figure 1b. The calculations are based on the recursive Green's function method and the tight-binding model with 10 hopping parameters extracted from DFT. Besides, the electron-phonon and electron-electron interactions are ignored. Furthermore, it is assumed that the device-lead coupling is perfect.

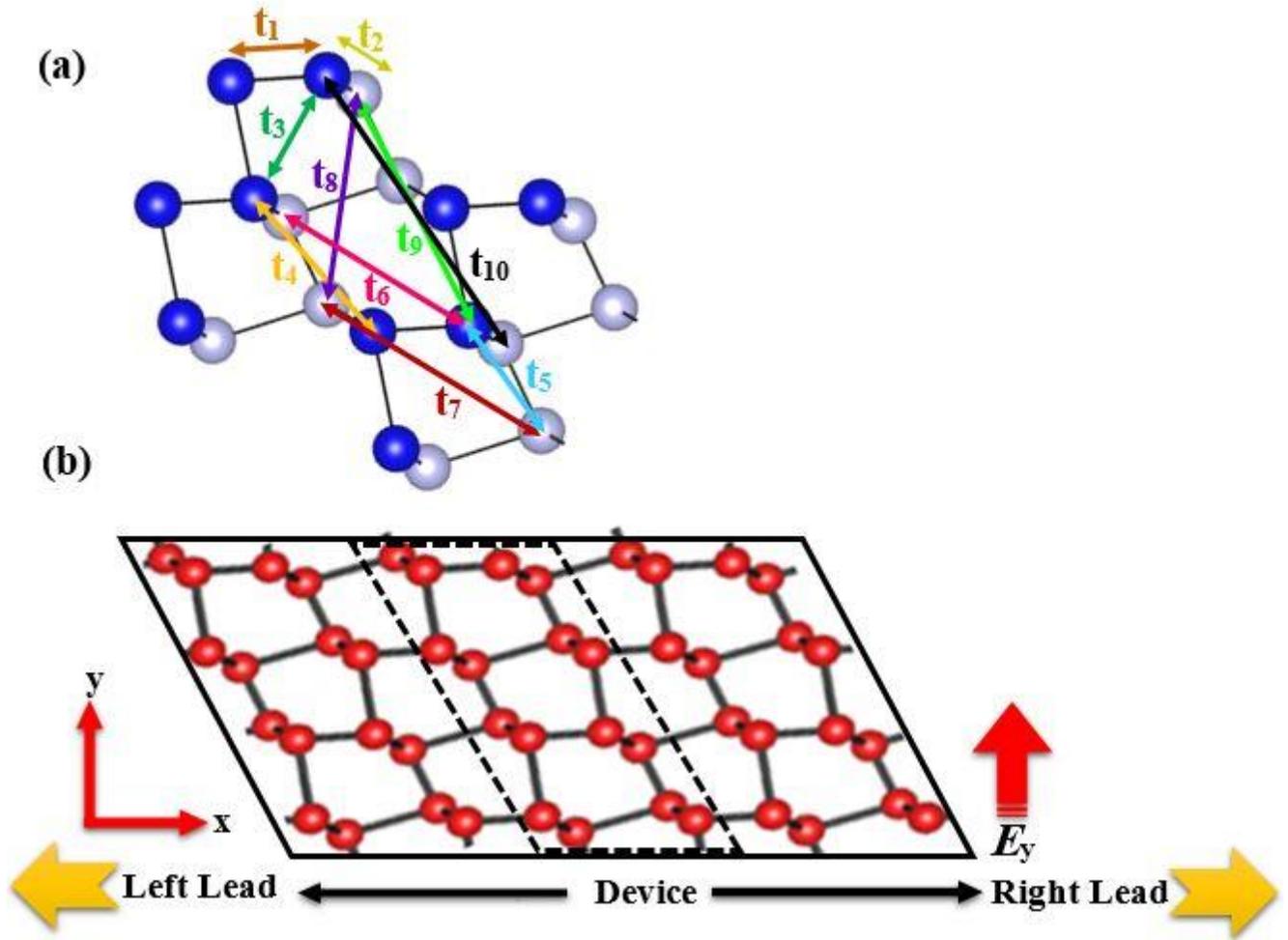

**Figure 1.** (a) A top view of a monolayer phosphorene including definition of 10 inter-atomic hopping energies. (b) Schematic of the skewed phosphorene nanoribbon composed of the left and the right leads. Here, the unit cell is introduced with dash box and $N_{sZZ}$ = 16 atoms. The unit cell is composed of two rows of atoms named even or odd in case of 80 or 81 atoms (in our calculations), respectively.

**Electronic Structure:** The easiest way to study the electronic properties of a pure system is to obtain its band structure. For this, we need the physical and geometric characteristics of the unit cells of the system. The unit cells of skewed-zigzag nanoribbons with $N_{sZZ}$=160 and 162 atoms in our study were generated using Visualization for Electronic and STructural Analysis (VESTA) [21]. There is a difference between the band structure of the phosphorene ribbon with widths (uni cells) including odd and even number of atoms as it will be explained later. The eigen values and eigenstates of each nanoribbon are obtained by diagonalizing the Bloch's Hamiltonian at each $k_x$ point:

$$H(k_x) = H_{00} + H_{-10}e^{(ik_x a_x)} + H_{-10}^\dagger e^{(-ik_x a_x)}, \qquad (1)$$

where is it assumed that the nanoribbon is periodic along x direction, and $H_{00}$ is intra-unit cell and $H_{-10}$ and $H_{-10}^\dagger$ correspond to inter-unit cell interaction Hamiltonians. $k_x$ represents the quasi-momentum (electron wave vector) along periodic *x* direction. The hopping parameters used in the above mentioned tight binding scheme (see figure 1a) were adapted from [17, 18] where the authors extracted the set of parameters by fitting the sub bands of bulk (two dimensional) phosphorene to that of DFT-based computations. The DFT calculations use partially self-consistent $GW_0$ approach with four maximally localized Wannier functions [22, 23].

During this work it was brought to our attention that M. Grujic et al. [12] used ten hopping parameters model [17] instead of using five parameters [16] and stated that the above model resolved finer details in the band structure.

**Density of states and conductance:** Electronic density of states is another important physical quantity for determining the electronic characteristics of a pure and disordered system. The density of states and conductance of the system were calculated using RGF method, in which effective Green's function of the device is given as [24]:

$$G(E) = ((E + i\eta)I - H_D - \Sigma_L(E) - \Sigma_R(E))^{-1}), \qquad (2)$$

where $G$ is the effective Green's function of the device and $E$ introduces the incident electron energy. Symbols L and R mean the left and the right leads, respectively and D for the device (figure 1b). $\Sigma_L$ and $\Sigma_R$ are self-energies for the left and the right leads, respectively. The system is described by a simple Hamiltonian in the nearest-neighbor tight-binding approximation as:

$$H = \sum_i \varepsilon_j C_i^\dagger C_i + e\frac{d}{2}E_z \sum_i \xi_j C_i^\dagger C_i + U + \sum_{<i.j>} t_{ij} C_i^\dagger C_j + H.C.. \qquad (3)$$

here $\varepsilon_j$ and $t_{ij}$ are on-site and hopping energies between the three of nearest-neighbor atoms, respectively. The on-site energies for the leads and the device are zero, $\varepsilon_L = \varepsilon_R = \varepsilon_D = 0$ eV, and the non-zero hopping energies are shown in figure 1a. $C_i^\dagger$ ($C_j$) is creation (annihilation) operator for electron on site $i$ ($j$) which varies between $(-\infty, +\infty)$. Here, $\xi = \pm 1$ and $d$ is layer thickness, also applied vertical electric field is introduced by $E_z$. $U = -eE_y y$ is the gate potential along the nanoribbon width which is added to the main diagonal of the Hamiltonian. $E_y$ is the applied transverse electric field. The transmission function of the system composed of the left lead, the right lead and the device can be calculated as:

$$T(E) = \text{Tr}\left(\boldsymbol{\Gamma}_L \boldsymbol{G}(E) \boldsymbol{\Gamma}_R \boldsymbol{G}^\dagger(E)\right), \quad (4)$$

in which $\boldsymbol{\Gamma}_{L(R)}$ is the broadening operator that is given by:

$$\boldsymbol{\Gamma}_{L(R)}(E) = i\left[\boldsymbol{\Sigma}_{L(R)}^\dagger(E) - \boldsymbol{\Sigma}_{L(R)}(E)\right]. \quad (5)$$

The self-energy operators $\boldsymbol{\Sigma}_L$ and $\boldsymbol{\Sigma}_R$ are calculated with the help of the RGF method [25-28]. These contain information about coupling of the device to the leads. The conductance of the system is evaluated at the zero temperature with the Landauer formula:

$$g = \frac{2e^2}{h} T(E). \quad (6)$$

Besides, the density of states at the energy $E$ is found from the imaginary part of the Green's function i.e.:

$$\text{DOS}(E) = -\frac{1}{\pi} \text{ImTr}\left(\boldsymbol{G}(E)\right). \quad (7)$$

**Charge Carrier Distribution:** To get an insight about the nature of the sub bands within the electronic structure, each eigenstate at each k point within the Brillouin Zone (BZ) is projected on its corresponding atom in the unit cell. As it is assumed that electron hopping (tunnelling) between the phosphorous atoms occurs only for $p_z$ orbitals, hence each number in the eigenstate corresponds to one orbital of each atom within the unit cell. In this case a Gaussian curve or a sphere is weighed according to the value of $|\Psi|^2$ and is centered on each atom to reflect the strength of $p_z$ contribution to that state.

**Effect of vertical (back gate) and transverse electric field:** By assigning the right amount of potential difference between the atoms which reside on the top and bottom layers of a single layer phosphorene, the vertical electric field can be simulated. Note that the distance between bottom and top atoms in a monolayer (along $z$ direction) is d = 2.14Å. The corresponding potential can be added directly to the main diagonal of the tight binding Hamiltonian. The transverse electric field results in a potential landscape in each unit cell according to $U = -eE_y y$. Hence the value of on-site energy of each atom is updated according to its $y$ coordinate.

**Optical dipole transition moment matrix element:** Within the dipole moment approximation the optical transition matrix element is written as the matrix element of position operator which is $\langle \psi_v | \boldsymbol{R} | \psi_c \rangle$. Where $\boldsymbol{R}$ is a diagonal matrix composed of $x$, $y$ or $z$ coordinates of all atoms within the unit cell. $\psi_{c,v}$ is the eigenstate corresponding to a conduction (c) or a valence (v) band, respectively.

### III. Results and Discussion

In this section, we demonstrate numerical results obtained from the effect of vertical and transverse electric field on electronic band structure and optical transition matrix elements for skewed-zigzag (sZZ) phosphorene nanoribbons. The investigated systems here are nanoribbons with 10 hopping parameters. They are built from 2×80 (160), 2×81 (162) atoms wide cells and are connected to two semi-infinite leads of the same type. By even and odd we refer to 80 and 81 atoms which compose half of a cell, respectively.

**Nature of flat bands for odd numbered width**: Due to the existence of various edge states and quantum size confinement in one dimensional nanoribbons, we expected a quasi-flat band dispersion near the Fermi level as mentioned in [12]. The deviation of mid-gap band from a perfect flat band is due to the highly electron and hole anisotropies in sZZ and sAC phosphorene nanoribbons [29]. The edge states are quasi-localized/delocalized on the outermost sites, which results in destructive/interactive interference of diffracted electron waves from edge atoms, so their amplitude exponentially decay into the bulk [30] and these nanoribbon-edge electrons can exhibit a pronounced splitting in their energy levels. The band structure, the density of states and the conductance of sZZ nanoribbons with even/odd number of atoms are plotted in figure 2. When $N_{sZZ}$ is an even number [i.e. 80 in figures 2a, b and c], we observed no edge mode (quasi-flat band) near the Fermi energy but in case of the unit cell with odd number of atoms [i.e. 81 in figures 2d, e and f], the mid-gap DOS and conductance show metallic characteristics. As will be seen later, application of transverse electric field creates the same metallic character. However, the degree of metallic behaviour depends on the slope of the mid-gap sub bands. If the mid-gap sub band has none-zero slope it may results in a non-zero group velocity for electrons. Otherwise the near zero slope of the sub band is more of a localized state which results in no charge movement.

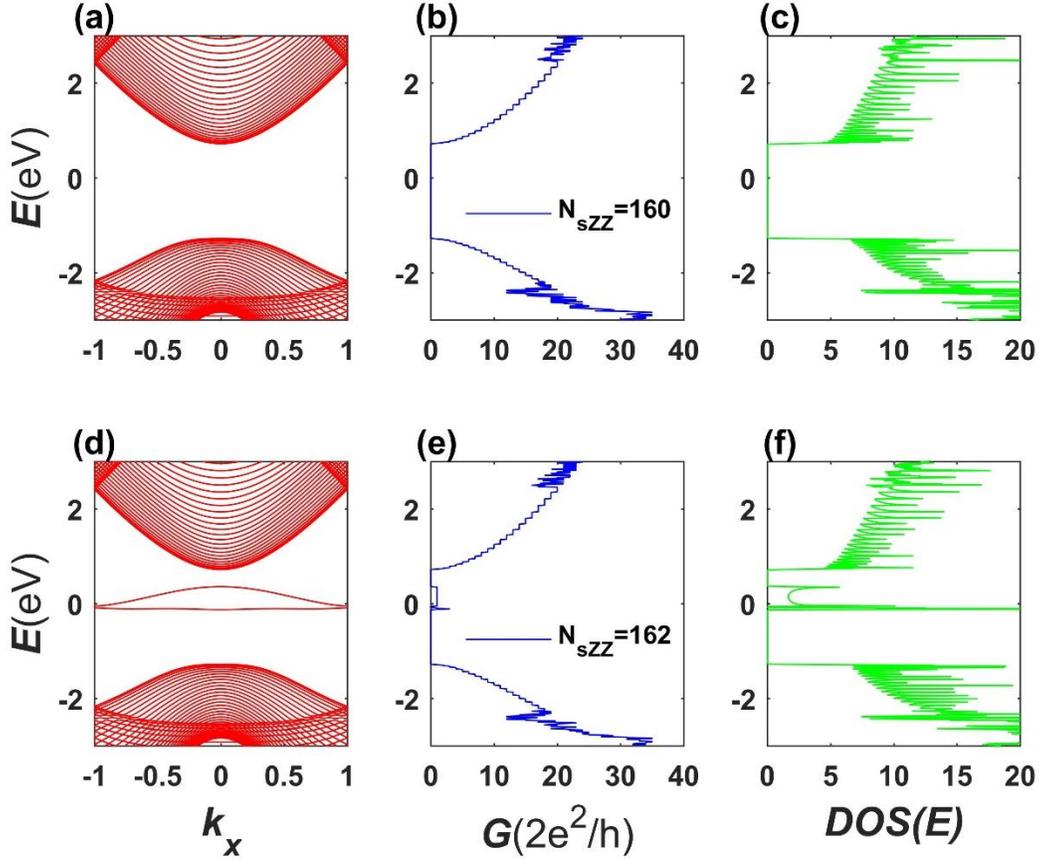

**Figure 2.** Band structure (red line), conductance (blue line) and the density of states (green line) for nanoribbons with even (80) number of atoms (a, b, and c). Panels d, e & f represent the band structure, the conductance and the density of states for odd (81) number of atoms.

**Optical dipole transition matrix element:** The optical dipole transition matrix element is a measure of optical transition between initial and final electronic states e.g. $|f\rangle$ and $|i\rangle$. The position and momentum of matrix element are related according to the following equation:

$$M_{fi} = \langle f|p|i\rangle = -i\frac{m_e}{\hbar}\langle f|[r,H]|i\rangle = im_e\omega_{fi}\langle f|r|i\rangle. \qquad (8)$$

where $m$ is mass of electron and $\omega_{fi}$ is frequency difference obtained from the bandgap or energy difference as $(E_i - E_f)/\hbar$. In order to have an optical transition e.g. a direct absorption of a photon, the dipole transition matrix element should be symmetry-allowed i.e. has a non-zero value. Hence symmetry of valence and conduction band state wave functions plays an important role. Here, the dipole matrix element for transitions between the Valence Band Maximum (VBM), the Conductance Band Minimum (CBM) and intermediate bands near the Fermi level (figure 3), were calculated at zero electric field. As can be seen for both even and odd numbered unit cells, the optical dipole transition between valence and conduction bands are of the same values. This suggests that the wave functions of valence and conduction bands in both nanoribbons have the same symmetry. Looking at figure 4 proves this as it shows the electron density distribution of valence and conduction states at the BZ center or $k_x = 0$ (in terms of $\pi/a_x$) on $x$-$y$ plane are of the same symmetry.

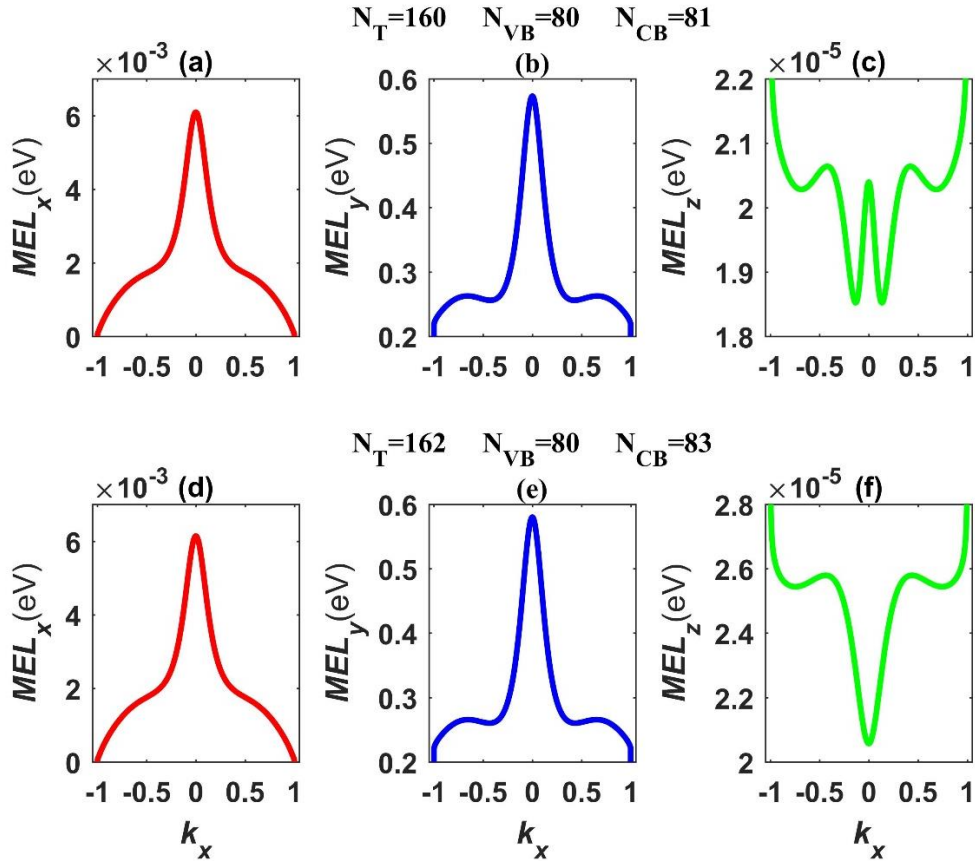

**Figure 3**. Optical dipole transition matrix element (MEL), in terms of (eV), between valence band (VB) and conduction band (CB) along 1D BZ of even numbered (2×80) nanoribbons at zero electric field. Panels a, b and c

represent the MEL values for *x*, *y* and *z*-polarized photons. Panels d, e and f show the same data for odd number of atoms (2×81).

It is also noteworthy, that matrix element for *z*-polarized photon is much smaller than that of *x* and *y* polarizations. This means that both *x*-polarized and *y*-polarized incoming photons with the suitable energy ($\hbar\omega \geq E_g$) can be absorbed better than a *z*-polarized photon. Additionally, the *y*-polarized photon has higher matrix element (about 0.6 eV) as opposed to *x*-polarized one (with maximum 0.006 eV at the BZ center or $k_x = 0$). This further supports this notion that when the polarization of the photon is along the width (*y*) or Armchair direction, there is higher interaction between the field and electronic distribution, hence larger matrix element is expected. Figure 4 also proves that the electron distribution is mostly elongated along *y* direction and increases the dipole interaction with electric field along *y* (transverse).

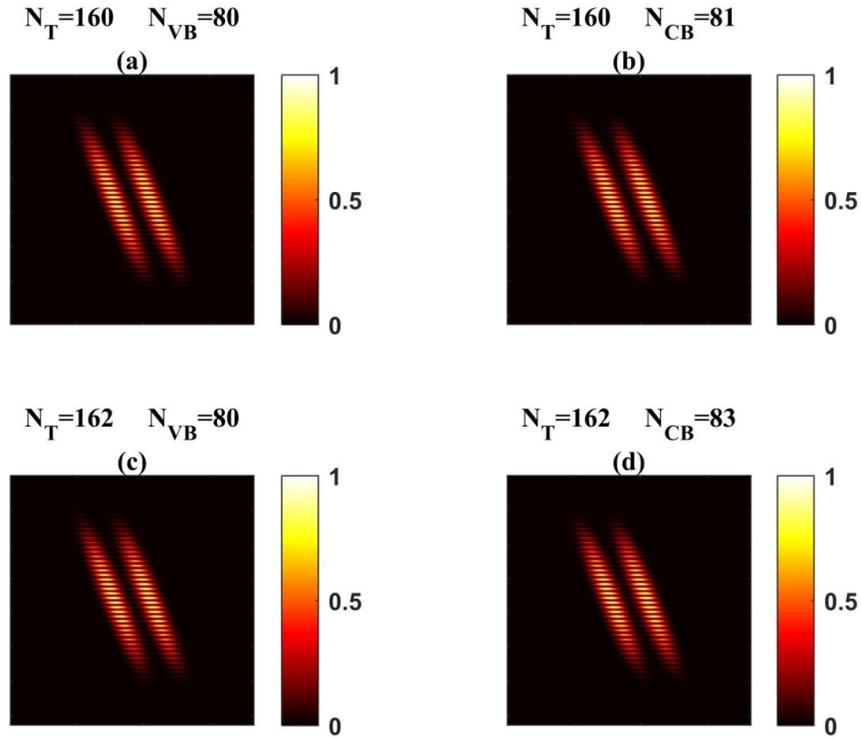

**Figure 4.** Electron density distribution ($|\psi|^2$) on the *xy* plane of the unit cell for valence (VB) and conduction (CB) bands at the BZ center ($k_x = 0$). Since for both nanoribbons, VB and CB have the same symmetry, MEL for x- and y-polarized photons are of the same value respect to $k_x$ (see figures 3a, b, d & e).

**Transverse electric field effect:** It was shown in [12] that an sZZ nanoribbon has a semiconducting band structure and an sAC nanoribbon has a metallic one. It should be noted that armchair (AC) nanoribbons have indirect bandgap and optical transitions are of phonon-mediated second order and very weak. In the case of sZZ nanoribbons, applying a transverse field along the ribbon width shrinks the band gap which renders it less interesting for Field Effect Transistors (FET) as $I_{on}/I_{off}$ ratio drops. The linear stark effect due to this field can lead to the splitting of degenerate energy levels. But for sZZ nanoribbons it conserves the degeneracy of quasi-flat bands and only the conduction (valence) bands shifts toward each other due to large potential difference in two edges. That results in the insulator-metal transition. Here, our system is exposed to a transverse gate voltage (e.g. $E_y$ = 30 mV/Å) along the width. In figure 5 we observe the band structure, the conductance and the DOS in the presence of $E_y$ which are in agreement with [12, 31]. It is observed that the transverse electric field causes upward and downward shift of energy for valence and conduction sub bands, respectively.

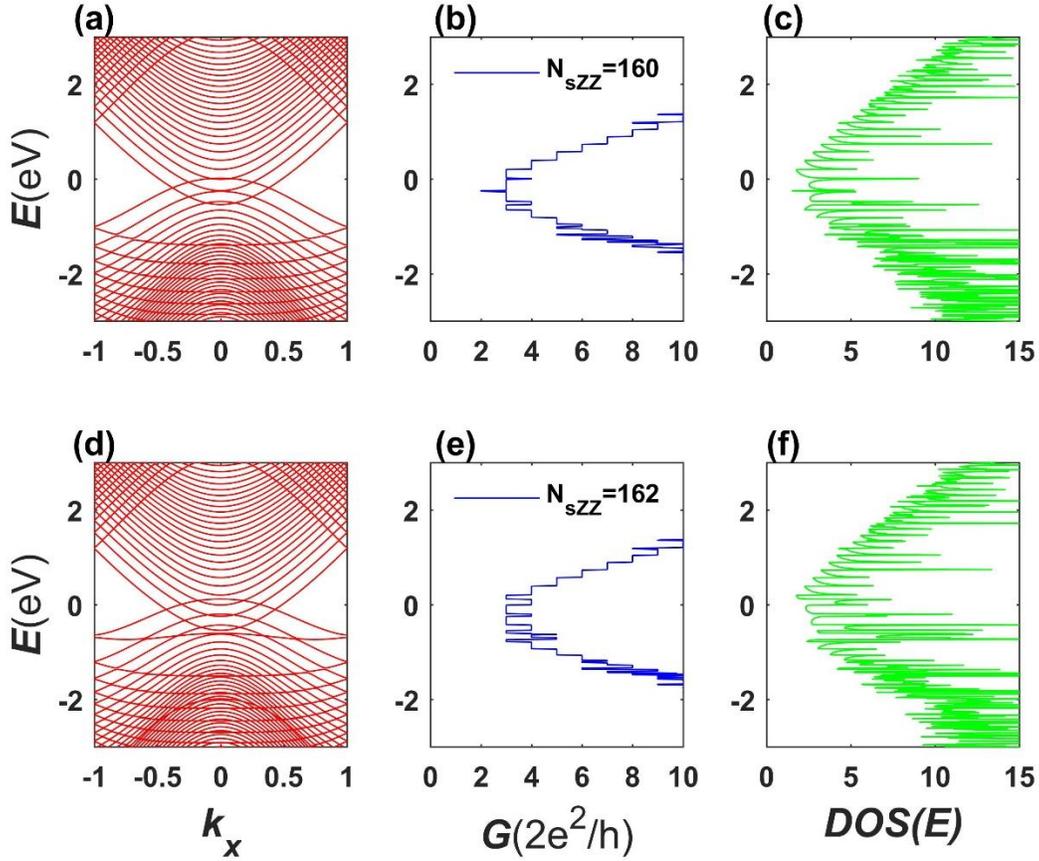

**Figure 5.** Down shift and up shift of conduction and valence sub bands in response to the transverse electric field ($E_y$=30 mV/Å) for 160 atoms (a, b and c), and 162 atoms (d, e, and f) in a unit cell. Red, blue and green colors represent the band structure, the conductance and the density of states.

**Effect of electric field (transverse and vertical) on matrix elements:** As it is shown in figure 6, by applying a transverse electric field in the direction of the *y*-axis ($E_y$), the matrix element value between the last VBM and the first CBM drops for all three photon polarizations (*x, y* and *z*). Also, this reduction in nanoribbons with odd number of atoms in width (see figures 6.d, e, and f), is more pronounced than even numbered unit cells close to the Fermi level. In fact, the transverse field shifts the valence bands upwards and the conduction downwards. This disturbance in the system is caused by changes in the structural symmetry and as a result, electron distribution i.e. emergence of localized states, at two edges of the ribbon. Figures 7a and b show the density distribution for valence and conduction bands at BZ center for even numbered (2×80 atoms) width. The same information for odd numbered width (2×81) is plotted at figures 7c and d for valence and conduction states at the BZ center, respectively. As it is evident electrons are pushed to the edge by the application of electric field along the width. The same scenario is true for holes but with reversed polarity i.e. accumulation of carriers on the bottom edge in response to the same field value and direction. Seeing this, it is understandable why transverse electric field reduces the MEL for even (figures 6a, b and c) and odd (figures 6d, e and f) atoms. The confinement of carriers on the edge significantly reduces the overlapping between conduction and valence wave functions which causes the MEL to vanish at the BZ center.

However the situation is different in the case of quasi-flat bands (QFB) and their corresponding edge states. Figure 8 shows the *x*- and *y*-polarized matrix elements for zero electric field (a and b) and $E_y$ = 30 mV/Å (c and d) for even number of atoms. The same data for odd number of atoms are shown in figures 8e and f (*x* and *y* polarizations, $E_y$ = 0) and figures 8g and h (*x* and *y* polarizations, $E_y$ = 30 mV/Å). Matrix elements correspond to the transitions from valance band to the QFB. The indices of the valence and QFB bands represent their corresponding number within the eigenstate (energy) spectrum. With regard to transitions like $|VB^{82}\rangle \rightarrow |QFB^{79}\rangle$ [figures 8c and d] and $|VB^{82}\rangle \rightarrow |QFB^{85}\rangle$ [figures 8g and h], the electric field enhances MEL values by one and two orders of magnitude, respectively. The zero electric field transitions i.e. $|VB^{80}\rangle \rightarrow |QFB^{81}\rangle$ figures 8a and b] and $|VB^{80}\rangle \rightarrow |QFB^{82}\rangle$ [figures 8e and f] have on the other hand a very small MEL.

Figures 9a and b reveals why MEL for VB to QFB transitions have very small values as the wave functions on *xy* plane have small overlapping. Increasing the transverse field on the other hand slightly enhances the overlapping (figures 9c and d).

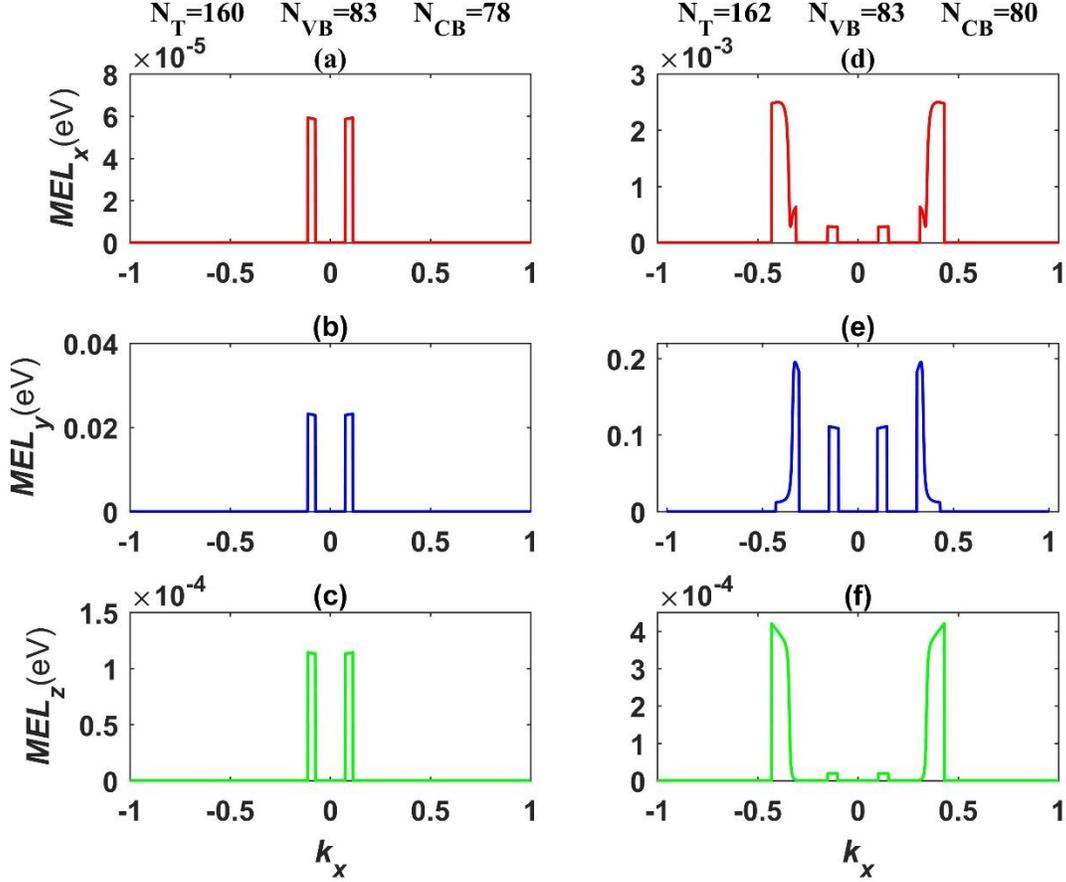

**Figure 6.** Effect of transverse electric field ($E_y$ = 30 mV/Å) on the band edge optical dipole transition matrix element. Red, blue and green represent *x*, *y* and z polarizations for photons. Panels a, b and c belong to the unit cell with even number of atoms (80). Panels d, e and f show the same for odd (81) number of atoms.

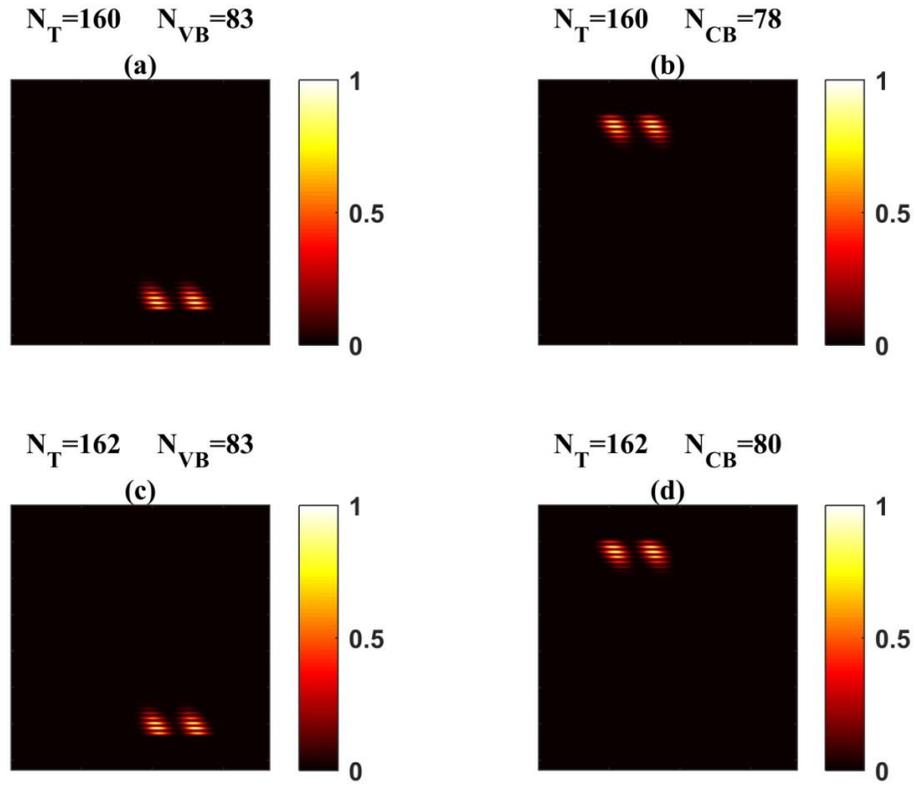

**Figure 7.** Localization of electron/hole distribution under the influence of transverse electric field ($E_y$ = 30 mV/Å). Panels a and b show the distribution of valence and conduction bands at BZ center for even (80) number of atoms. Panels c and d show the same data for odd (81) number of atoms.

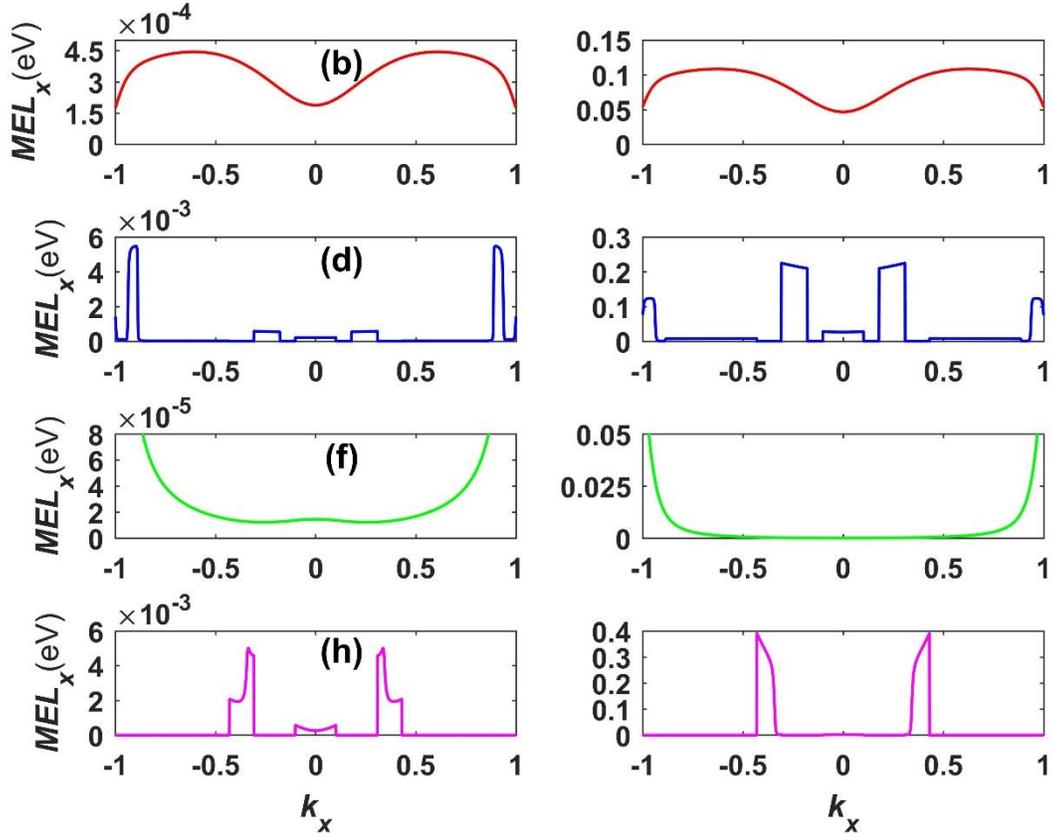

**Figure 8.** Left and right panels show the MEL values for *x* and *y* polarization of photons, respectively. Under $E_y = 30$ mV/Å, the localized states (QFBs 79 and 85) have non zero overlapping with VB82 (as shown in c, d, g, and h) which leads to the enhancement of MEL as opposed to (a, b, e, and f) in which $E_y = 0$ and transition occurs between QFBs 81 and 82 with VB80.

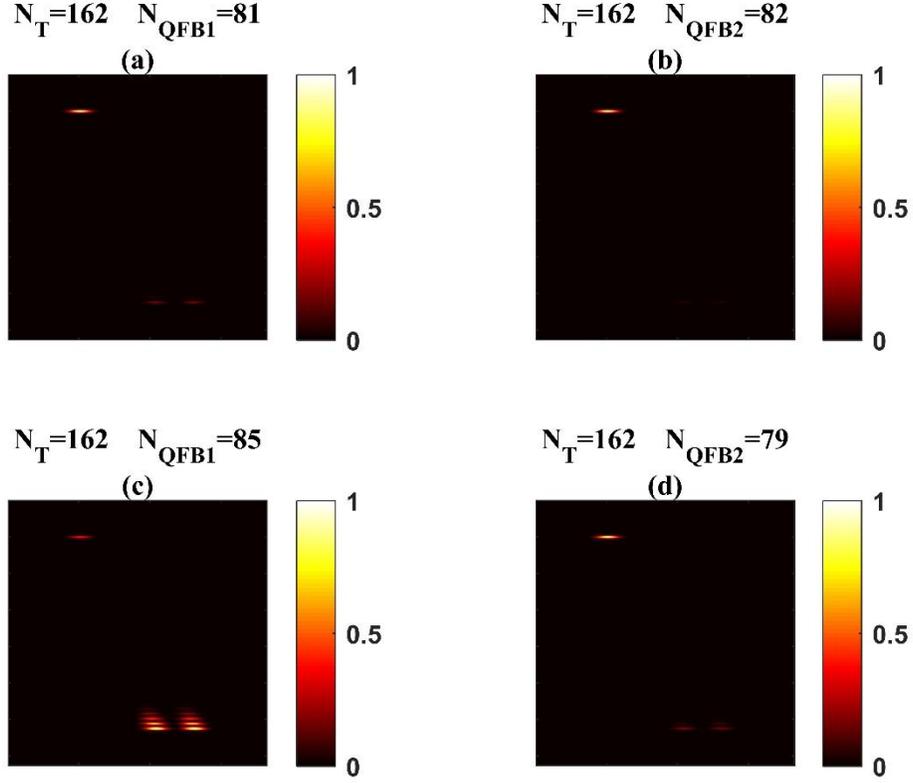

**Figure 9.** The electron distribution of edge states near the Fermi level for odd numbered unit cell (81 atoms) in with $E_y = 0$ (a, b) and $E_y = 30$ mV/Å (c, d). Nonzero overlapping of QFB with VB leads to enhanced matrix element as shown in figures 8, c, d, g, and h.

**Vertical electric field effect:** In contrast to the normal nanoribbons, in the case of sZZ nanoribbons, the vertical electric field lifts the degeneracy of all bands including QFBs. The conduction and valence bands are shifted to opposite directions in energy. Figure 10 shows the band structure, the conductance and the density of states for even (a, b and c) and odd (d, e, an f) number of atoms at $E_z = 400$ mV/Å.

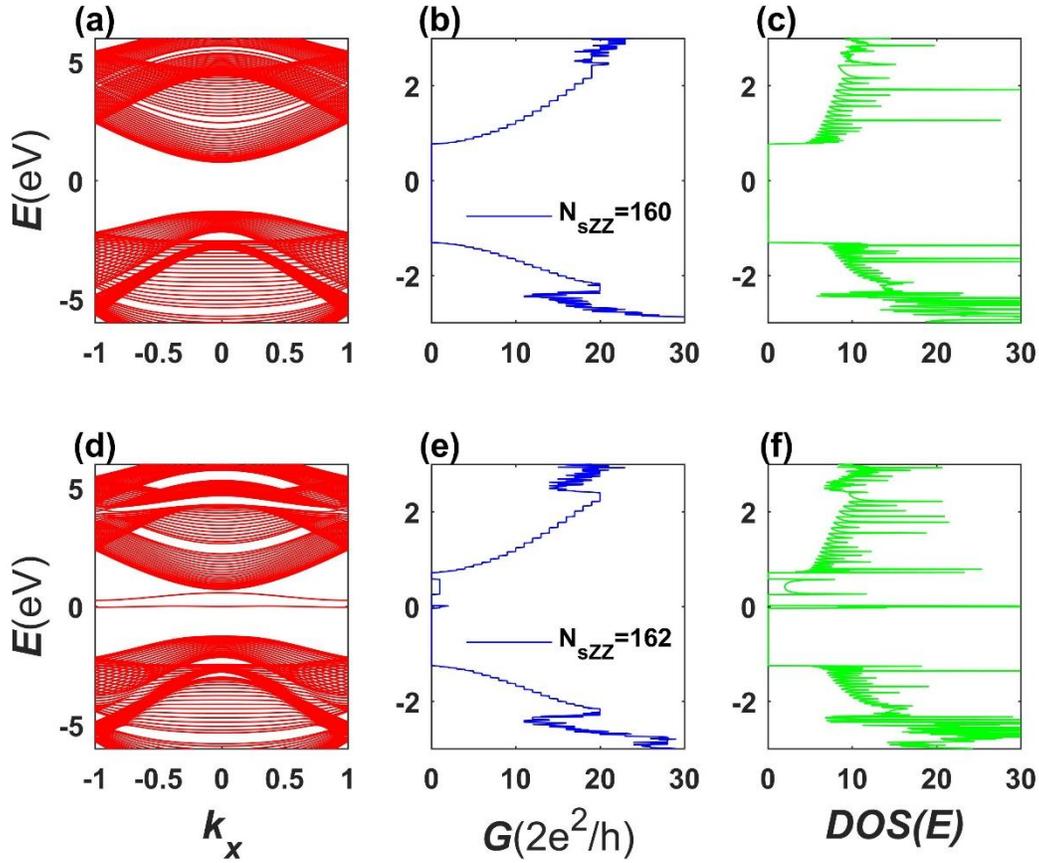

**Figure 10.** Effect of vertical electric field, $E_z$ = 400 mV/Å on even (top row) and odd numbered (bottom row) unit cells ($N_{sZZ}$= 160 and 162). Red, blue and green panels represent the band structure, the conductance and the density of states, respectively.

To simulate the effect of vertical electric field, suitable values of potential energy is added to main diagonal of the Hamiltonian for corresponding atoms on top and bottom planes (along *z* direction). The nature of mid-gap sub bands for $N_{sZZ}$ = 162 can be understood by looking at the band structure as shown in figure 10d. It is also noteworthy that upon application of the field, a sub band bunching effect occurs. This can be understood be recalling figure 1a, in which there are two atoms on higher *xy* plane (*z* = + 1.07Å) and two atoms on lower *xy* plane (*z* = - 1.07Å). Since half of the atoms in each unit cell are feeling a higher potential, then their corresponding sub bands in the BZ will move higher in energy, in contrast half of the atoms feel a lower potential, hence their corresponding sub bands move lower in energy. Now as it is evident in figures 10a and d, half of those up shifting atoms compose the valence sub bands and half of them compose the conductions sub bands likewise. The same is true for up shifting (lower plane) atoms.

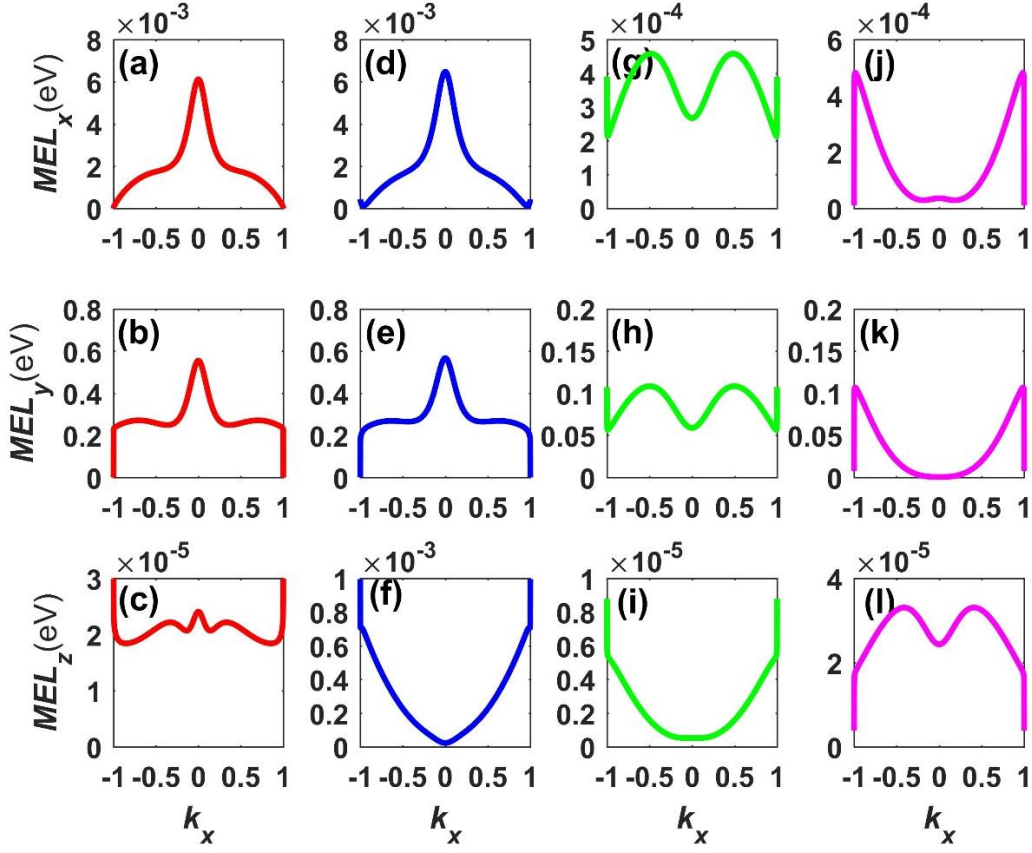

**Figure 11.** Effect of vertical electric field, $E_z = 400$ mV/Å, on the band edge optical dipole transition matrix element plotted in terms of $k_x$ for even width (a, b, c), and odd width (d, e, f). Corresponding data for transitions from the valence bands to the first (g, h, i) and the second (j, k, l) quasi-flat bands in odd numbered width. The first, second and third row represent $x$, $y$ and $z$ polarizations.

Comparing figure 3 with figures 11a,b,c and figures 11d,e,f suggests that, the matrix element values do not show any significant change by applying vertical electric field up to $E_z = 400$ mV/Å. This is because the symmetry of wave functions or density distributions are similar to the cases without transverse electric field (compare figure 4 and figures 12a & b). Also, we observed that the transition to quasi-flat bands is extremely low, because their $|\psi|^2$ distribution is highly localized and as a result the overlapping integral in $\langle \psi_v | r | \psi_c \rangle$ is very small.

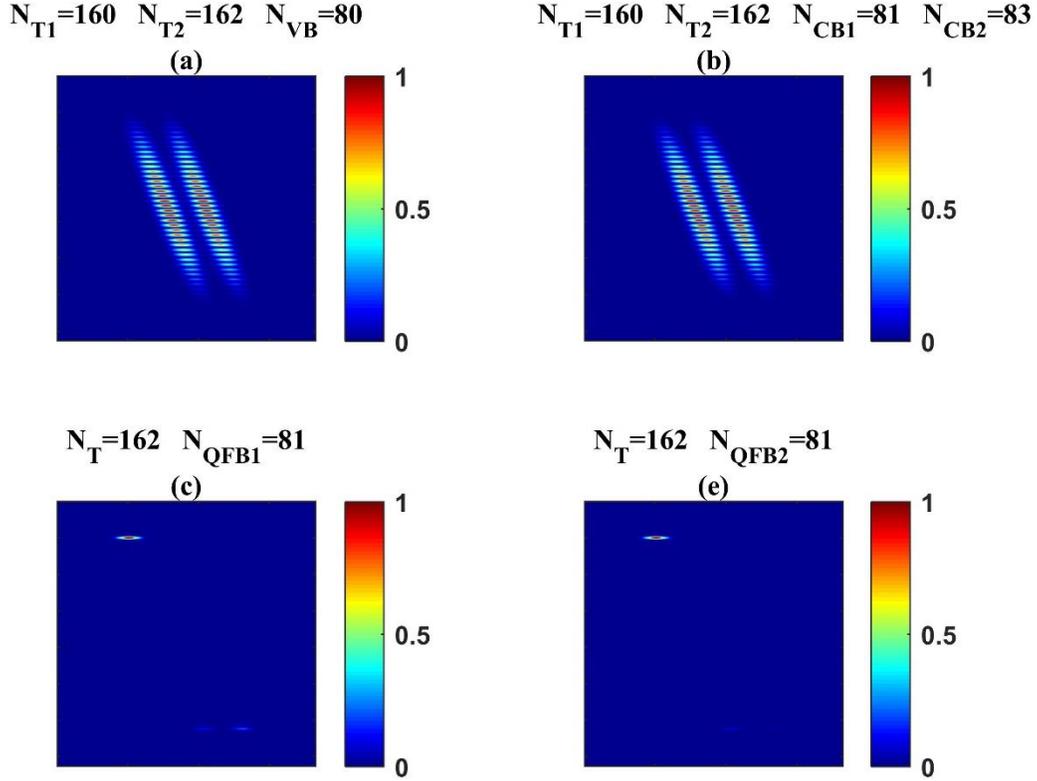

**Figure 12**. Eelectron/hole distribution under the influence of a vertical electric field ($E_z = 400$ mV/Å) in an odd/even numbered width (80, 81atoms) for VB and CB (a, b). The same data for QFB1 and QFB2 (c and e) in an even numbered unit cell.

### IV. Conclusions

We observed that by applying *transverse electric field* to the skewed-zigzag (sZZ) phosphorene nanoribbons: (1) the two-fold degeneracy of quasi-flat bands can be conserved and this causes a semiconductor-to-metal transition, (2) it decreases the transition matrix element between the last valence band (VBM) and the first conduction band (CBM) in both cases of odd and even width unit cells, but it can enhance the matrix element for intermediate transitions (QFBs) in even numbered widths. (3) Additionally, the electron density distributions reveal localized nature of two sub bands in the latter case.

If the nanoribbon is subjected to the *vertical electric field*, (1) no significant change in matrix elements was observed for moderate values of the field. (2) However there is a lift of degeneracy observed for QFBs. Additionally (3) a bunching of valence and conduction sub bands emerges in response to the vertical field. It should be noted that transition to QFBs leads to the electron localization on both sides of ribbon. These effects are of significance for electronic and optical devices based on phosphorene e.g. voltage controlled photo detector. Large bandgap as opposed to pristine graphene provides opportunities for phosphorene in active electronic devices as well.